# Three dimensional Dirac point at $\vec{k}=0$ in photonic and phononic systems

Xueqin Huang, Fengming Liu, and C. T. Chan[*]

*Department of Physics, Hong Kong University of Science and Technology, Clear Water Bay, Kowloon, Hong Kong, China*

*Corresponding author: phchan@ust.hk

**Abstract**

While "Dirac cone" dispersions can only be meaningfully defined in two dimensional (2D) systems, the notion of a Dirac point can be extended to three dimensional (3D) classical wave systems. We show that a simple cubic photonic crystal composing of core-shell spheres exhibits a 3D Dirac point at the center of the Brillouin zone at a finite frequency. Using effective medium theory, we can map our structure to a zero refractive index material in which the effective permittivity and permeability are simultaneously zero at the Dirac point frequency. The Dirac point is six-fold degenerate and is formed by the accidental degeneracy of electric dipole and magnetic dipole excitations, each with three degrees of freedom. We found that 3D Dirac points at $\vec{k}=0$ can also be found in simple cubic acoustic wave crystals, but different from the case in the photonic system, the 3D Dirac point in acoustic wave system is four-fold degenerate, and is formed by the accidental degeneracy of dipole and monopole excitations. Using effective medium theory, we can also describe this acoustic wave system as a material which has both effective mass density and reciprocal of bulk modulus equal to zero at the Dirac point frequency. For both the photonic and phononic systems, a subset of the bands has linear dispersions near the zone center, and they give rise to equi-frequency surfaces that are spheres with radii proportional to $(\omega - \omega_{Dirac})$.

## I. Introduction

Many interesting physical properties of graphene are consequences of the Dirac cone dispersion in its electronic band structure [1-5]. It turns out that some classical wave systems such as 2D photonic and phononic crystals also possess Dirac cone dispersions [6-19]. Intriguing transport properties such as "Zitterbewegung" [17-20] and some interesting physical properties such as the possession of non-trivial Berry phases [10-15] that originate from the Dirac cone in electronic systems can also been observed in classical wave systems possessing Dirac cone dispersions. Most of the Dirac points, including those in graphene [1-5] and 2D photonic/phononic crystals [9-14, 16-19], are found at the Brillouin zone boundary. However, it was recently discovered that Dirac points can also be realized at the zone center of 2D photonic [6] and phononic [7-8, 15] crystals. The physical origins of these $\vec{k} = 0$ Dirac-like cones are different from the usual zone boundary Dirac cones. Whereas zone boundary Dirac cones are typically consequences of lattice symmetry, the existence of Dirac points at $\vec{k} = 0$ requires some sort of accidental degeneracy. Zone center Dirac-like cones are also accompanied by an extra flat band of states [6-8, 15], which makes the physics different from that of zone boundary Dirac cones. For example, the Berry phase enclosing a zone center Dirac point is zero [15]. In a previous paper, we found that a subset of such photonic crystals behave as if they have effectively *ε=0* and *μ=0* at the Dirac point frequency [6], and the analogy can be extended to acoustic [7] and elastic [8] waves. So, 2D photonic and phononic crystals with Dirac points at $\Gamma$ ($\vec{k}=0$) can be described as zero-refractive-index materials under certain conditions [6-8]. Recently, Sakoda showed that it is possible to find 3D Dirac points at the Brillouin zone center of cubic electromagnetic systems with $O_h$ symmetry [21]. According to Sakoda, a 3D Dirac point can be created by the accidental degeneracy of a non-degenerate $A_{1g}$ mode, and a triply-degenerate $T_{1u}$ mode in 3D photonic systems [21]. The concept of Dirac point is hence generalized from 2D to 3D. We should note here that the notion of a Dirac "cone" is inherently a 2D concept. To make our discussion more straightforwardly, we will limit ourselves to isotropic systems in the following. The equi-frequency surfaces corresponding to 2D Dirac cones are elements of a set of circles whose radii decrease linearly and approach zero both from above and below the Dirac point and the Dirac point is the frequency at which the equi-frequency circle has zero radius. To obtain a Dirac cone, the dispersion must be linear near the zone center which requires accidental degeneracy. When concept is generalized from 2D to 3D, the equi-frequency trajectories change from circles in 2D to spheres in 3D. Near the "Dirac point" in 3D systems, the radii of equi-frequency spheres are linearly proportional to $(\omega - \omega_{Dirac})$ where $\omega_{Dirac}$ is the Dirac point frequency, and the Dirac point is the frequency at which the equi-frequency sphere becomes a point. The necessary condition is again linear dispersions near the zone center which requires accidental degeneracy.

While Sakoda has already shown mathematically that it is possible to extend the Dirac point concept to 3D and has shown using group theoretic method that such 3D Dirac point can be obtained using a singly-degenerate $A_{1g}$ mode and a triply-degenerate $T_{1u}$ mode in a simple cubic system [21], there is still more work to do. We wish to explore the following issues in this paper. Firstly, we will explore whether some of these 3D Dirac point systems can be mapped to a system with $\varepsilon_{eff} = \mu_{eff} = 0$ at the Dirac point, as in the case of 2D systems [6]. In addition, we want to explore whether there are simple physical systems that can realize such system with accidental degeneracy. For realization in physical systems, we note that the $A_{1g}$ mode correspond to the monopolar excitations. This poses a challenge in practice because monopoles are difficult to construct in 3D photonic systems. The lowest order excitations are typically dipoles. We will show that the accidental degeneracy of 3D dipoles and monopoles can be realized instead in acoustic wave systems, and thereby obtaining Dirac points in 3D for acoustic crystals. For electromagnetic waves in 3D, it is more challenging but we will show that there is another route to obtain a Dirac point. Using group theoretic approach and by giving a specific example, we will show that 3D Dirac points can be obtained if electric and magnetic dipole excitations can be arranged to be accidentally degenerate at $\vec{k}=0$.

More specifically, we demonstrate that core-shell spheres arranged in a simple cubic lattice exhibit a 3D Dirac point at the Γ point. The Dirac point consists of six-fold-degenerate states, formed by the triply-degenerate $T_{1u}$ and $T_{1g}$ modes. The $T_{1u}$ and $T_{1g}$ modes correspond to magnetic and electric dipoles respectively. Four of the six-fold-degenerate states have linear dispersions near the Dirac point, while the other two are quadratic. This Dirac point is different from the scheme proposed in Ref. [21], which is a four-fold-degenerate state. For acoustic wave systems, we found that a simple cubic system composing of rubber spheres embedded in the water can give a 3D Dirac point at the Γ point in the band structure and the Dirac point is formed by a four-fold-degenerate state. This four-fold-degenerate state comprises a non-degenerate monopole $A_{1g}$ mode and a triply-degenerate dipole $T_{1u}$ mode, which falls into the class of 3D Dirac point proposed in Ref. [21]. For the 3D acoustic wave system, the tight-binding method shown in Ref. [21] can be applied directly to prove the existence of the Dirac point. The 3D Dirac point in photonic systems has six degrees of freedom, and as such, the proof of the existence of the Dirac point is more tedious and the details will be shown in the appendix.

Using effective medium theory, we will show that the structures with 3D Dirac point in photonic and phononic systems can be described as isotropic zero-refractive-index materials at the Dirac point frequency, in the sense that all components of the effective permittivity and permeability are equal to zero in the photonic system, and the effective mass density and reciprocal of bulk modulus are equal to zero in the phononic system.

## II. 3D Dirac point in photonic system

It is well known that photonic bands have quadratic dispersions at the Γ point as a consequence of time reversal symmetry. In order to illustrate the typical dispersion of a simple cubic lattice near the zone center, we calculate the photonic band structure of dielectric spheres arranged in a simple cubic lattice, which is shown in Fig. 1. The permittivity and permeability of the spheres are chosen to be *ε=12* and *μ=1*. The radius is *R=0.3a*, with *a* being the lattice constant. In the frequency range of interest, the bands are derived from electric and magnetic dipolar excitations. There are two triply-degenerate states at Γ point and these states have different frequencies. The lower bands correspond to magnetic dipole modes which are labeled as $T_{1u}$ at $\vec{k}=0$, and the higher bands are the electric dipole modes labeled as $T_{1g}$ at $\vec{k}=0$ [22]. Both sets of the bands have quadratic dispersions at the Γ point as expected. We cannot obtain linear dispersion at Γ point unless there is accidental degeneracy. However, in photonic crystals composing of the dielectric spheres arranged in a simple cubic lattice, it is very difficult, if not impossible, to make the $T_{1u}$ and $T_{1g}$ modes touch each other at the same frequency through tuning the filling ratio or permittivity of the spheres. To achieve accidental degeneracy, we need one more freedom to tune the frequencies of these two modes. For that purpose, we employ a core-shell structure. The inset in Fig. 2(a) is the illustration of the simple cubic unit cell. The core (orange color in the figure) is a perfect electric conductor, the radius of which is *$R_1$=0.102a*.

The shell (gray color in the figure) is dielectric with $\varepsilon=12$ and $\mu=1$. The outer radius of the shell is $R_2=0.3a$. The photonic band structure of these core-shell spheres arranged in a simple cubic lattice is shown in Fig. 2(a). With the extra degree of freedom allowed in a core-shell configuration, it is not difficult to find parameters so that the $T_{1u}$ and $T_{1g}$ modes touch each other at $\Gamma$ point. For the system shown in Fig. 2, the accidental degeneracy occurs at the frequency $f_{Dirac}=0.523c/a$. As the electric and magnetic dipoles are each three-fold degenerate at $\vec{k}=0$, the accidental degeneracy gives rise to a six-fold degenerate state at $\Gamma$ point. Four bands have linear dispersions near the $\Gamma$ point, the other two bands are relatively flat.

To visualize the accidental degeneracy further, we calculate the band structure of the core-shell spheres with different radii of the core. The outer radius of the shell is kept constant ($R_2=0.3a$). The photonic band structures for $R_1=0.08a$ and $R_1=0.12a$ are shown in Fig. 3(a) and 3(b), respectively. The six-fold degenerate state (shown in Fig. 2(a)) breaks into two triply-degenerate states. When $R_1$ is smaller than $0.102a$, the frequency of $T_{1u}$ mode is lower than that of $T_{1g}$ mode. If $R_1$ is bigger than $0.102a$, $T_{1u}$ mode becomes the higher frequency mode. It is quite obvious that by changing the radius of the core, the $T_{1u}$ and $T_{1g}$ modes will cross each other at the same frequency, which is consistent with the result shown in Fig. 2(a). Figure 3 also shows that unless $T_{1u}$ and $T_{1g}$ modes coincide in frequency, the dispersion remains quadratic as in the case of simple dielectric spheres shown in Fig. 1.

To prove the existence of the linear dispersion near the $\Gamma$ point, we extend the method introduced by Sakoda in Ref. [21] to the more complicated case of six-fold degeneracy. We will use the same notation introduced in Ref. [21]. The secular equation of this 3D accidental-degeneracy-induced Dirac point (ADIDP) is:

$$\left| \mathbf{B} - \frac{\omega_k^2}{c^2}\mathbf{I} \right| = 0. \qquad (1)$$

Here, **I** is the unit matrix. $\omega_k$ is the eigen frequency. The elements ($B_{ij}$) of **B** is defined by:

$$B_{ij} = \sum_{lmn} e^{ia(k_x l + k_y m + k_z n)} L_{lmn}^{(ij)} \qquad (2)$$

Here, *l, m, n* are integers, $k_x$, $k_y$, $k_z$ are the Cartesian components of the Bloch wave vector and $L_{lmn}^{(ij)}$ are the "electromagnetic transfer integrals" in 3D. By substituting the expressions of $L_{lmn}^{(ij)}$ into Eq. (2), we can obtain the formulas of $B_{ij}$, which allows us to solve Eq. (1). The detailed derivations of the secular equation, $L_{lmn}^{(ij)}$, and $B_{ij}$ are shown in the appendix. After some tedious calculations, we can solve Eq. (1) to obtain the dispersion of the 3D ADIDP.

$$\omega_k = \begin{cases} \omega_\Gamma + |M_3| ac^2 k / \omega_\Gamma - \left(\frac{1}{12}(M_1' + 2M_1'' + M_2' + 2M_2'')/\omega_\Gamma - \frac{1}{2\omega_\Gamma^3} c^2 |M_3|^2\right) a^2 c^2 k^2 & (double\ roots) \\ \omega_\Gamma - |M_3| ac^2 k / \omega_\Gamma - \left(\frac{1}{12}(M_1' + 2M_1'' + M_2' + 2M_2'')/\omega_\Gamma - \frac{1}{2\omega_\Gamma^3} c^2 |M_3|^2\right) a^2 c^2 k^2 & (double\ roots) \\ \omega_\Gamma - \frac{1}{12}(M_1' + 2M_1'' + M_2' + 2M_2'') a^2 c^2 k^2 / \omega_\Gamma & (double\ roots) \end{cases} \qquad (3).$$

Equation (3) implies that in the band structure four bands have linear dispersions near the zone center, and the other two are quadratic in *k* in the lowest order. The dispersions are isotropic near $\vec{k} = 0$. This is consistent with the numerical results shown in the band structure in Fig. 2(a). The equi-frequency surfaces for the linear bands are spheres with radii proportional to the $(\omega - \omega_{Dirac})$.

In order to understand the underlying physics of this 3D ADIDP, we calculate the electric field patterns of the eigenstates at the Dirac point frequency (*0.523c/a*). Figure 4(a) demonstrates the electric field pattern in the *yz* plane (*x=0* plane). The arrows show the directions of the electric field, the sizes of the arrows show the amplitude of the field. The field pattern shows that this eigenstate is a magnetic dipole pointing along the *x* direction. Figure 4(b) shows the electric field pattern in the *xy* plane (*z=0* plane), which looks like an electric dipole. The eigenstates of the other four modes at the Dirac point frequency correspond to magnetic and electric dipole excitations along *y* and *z* directions (not shown here). Therefore, the 3D ADIDP can be described by electric and magnetic dipole excitations along *x, y,* and *z* directions.

We have shown previously that in 2D photonic crystals, Dirac points that are derived from the accidentally degeneracy of monopole and dipole excitations can be mapped to a zero-refractive-index at that frequency [6]. Is it possible to describe the Dirac point physics in 3D with effective medium theory? We apply the effective medium theory shown in Ref. [23] to calculate the effective medium of this 3D ADIDP. The effective permittivity ($\varepsilon_{eff}$) and permeability ($\mu_{eff}$) as a function of frequency are shown in Fig. 2(b). At the 3D Dirac point frequency, effective medium theory found that $\varepsilon_{eff} = \mu_{eff} = 0$. That is to say, the core-shell simple cubic structure will behave as if it is an isotropic system with $\varepsilon_{eff} = \mu_{eff} = 0$ at the Dirac point frequency. This is different from the case of 2D Dirac cone systems which are anisotropic and as such, can only give zero refractive index only for one polarization.

Zero-index material has many interesting wave manipulating properties [24-35]. We perform finite-difference time-domain simulations to see whether our core-shell structure has properties similar to a zero-index material. Figure 5 shows a snapshot of field patterns of incident plane waves going through the core-shell photonic crystal. Due to the limitation of computation resources, we use a relatively small super cell to do the simulations (shown in Fig. 5(a)). The side walls (*xz*, and *yz* planes) have periodical boundary conditions. There is only one unit cell in the *x* direction, while there are $N_y$ unit cells in the *y* direction. Perfectly matched layer boundary conditions are imposed on the top and bottom walls (*xy* planes). The sample is $N_z$ unit cells thick in the *z* direction. In our simulations shown, we use $N_y=4$, $N_z=6$. The core-shell spheres are arranged from *-1.5a* to *1.5a* and *-2.5a* to *2.5a* in the *y* and *z* directions. The incident plane waves come in from the negative *z* direction as indicated by the pink arrows in the figure with the polarization of the electric field along the *x* direction (blue arrows). In Fig. 5(b), the electric field pattern in the *xz* plane (*y=0* plane) preserves its plane wave front without distortion. In *yz* plane (*x=0* plane), the electric field pattern also preserves its plane wave front (shown in Fig. 5(c)). Examining the field patterns inside the core-shell photonic crystal (shown in Fig. 5(b) and 5(c)), we can find little phase inside the photonic crystal, which is expected if the material has $\varepsilon_{eff} = \mu_{eff} = 0$.

## III. 3D Dirac point in phononic system

From the previous discussions, we see that 3D ADIDP in photonic systems can be achieved with the accidental degeneracy of electric and magnetic dipoles resulting in a six-fold degenerate Dirac point at the zone center. We are going to show below that the scheme proposed in Ref. [21], using $A_{1g}$ (one-fold) and $T_{1u}$ modes (three-fold) to get 3D Dirac point, can actually be employed in acoustic wave systems rather than in photonic systems. We calculate the band structure of 3D simple cubic lattice acoustic crystals consisting of rubber spheres in water. The radius of rubber is $R=0.255a$, with $a$ being the lattice constant. The density of rubber is taken to be $\rho = 1.3 \times 10^3 \, kg/m^3$, and that of water is $\rho_0 = 1.0 \times 10^3 \, kg/m^3$. The Lame constant in rubber is taken to be $\kappa = 1.17 \times 10^8 \, N/m^2$ and for water $\kappa_0 = 2.22 \times 10^9 \, N/m^2$. For simplicity, we have ignored the shear wave within the rubber spheres due to the high velocity contrast between the rubber and water, and the main features will stay the same if we also include the shear wave within the spheres [36]. Figure 6(a) shows a four-fold degenerate state at the Γ point at a frequency $f_{Dirac}=0.418v_0/a$. Here, $v_0$ is the acoustic velocity of water. Two bands have linear dispersions near $\vec{k} = 0$, and the other two have quadratic dispersions. For the linear bands, the equi-frequency surfaces are spheres with radii proportional to $(\omega - \omega_{Dirac})$. This Dirac point is also induced by accidental degeneracy. However, it is obviously different from the 3D ADIDP in the photonic system, which is six-fold degenerate. To visualize the accidental degeneracy further, we calculate the band structure of the acoustic crystal with different radii (shown in Fig. 7) of the rubber spheres. The four-fold degenerate state now breaks into a singly-degenerate state ($A_{1g}$ mode) and a triply-degenerate state ($T_{1u}$ mode), both with zero group velocity at the zone center. When the radius is smaller than *0.255a* (the radius satisfies the 3D Dirac point condition), the frequency of the singly-degenerate state is lower than the triply-degenerate state (Fig. 7(a)). If the radius is bigger than *0.255a*, the ordering reverses (Fig. 7(b)). The ordering of the states for spheres with different radii implies that at certain radius, these two modes will touch each other at the same frequency, which is consistent with the results shown in Fig. 6(a). To understand the underlying physics, we calculate the displacement field pattern of the eigenstates at the Dirac point frequency shown in Fig. 8. The displacement patterns in Figs. 8(a) and 8(b) are consistent with a monopole and dipole excitation, respectively. The monopole and dipole modes correspond to the $A_{1g}$ and $T_{1u}$ modes. The existence of the 3D ADIDP in this acoustic crystal with four degrees of freedom can be proved with the method shown in Ref. [21].

Using effective medium theory [37], we can calculate the effective mass density ($\rho_{eff}$) and reciprocal of bulk modulus ($1/\kappa_{eff}$) of this acoustic crystal. The effective mass density and reciprocal of bulk modulus are simultaneously equal to zero at the frequency *0.418$v_0$/a*. This frequency is the same as the Dirac point frequency in the band structure. We note that in 2D geometries, acoustic systems with Dirac point derived from the accidental degeneracy of monopole and dipole excitations can be mapped to a zero-index effective medium [7]. The results here show that the same analogy exists for Dirac points in 3D derived from $A_{1g}$ and $T_{1u}$ degeneracy.

## IV. Summary

In conclusion, we proposed physically realizable structures to obtain 3D ADIDP in both photonic and phononic systems. The Dirac point in 3D photonic systems is a six-fold degenerate state, formed by the accidental degeneracy of electric and magnetic dipoles excitations, while in the acoustic wave system it is a four-fold degenerate state, formed by the monopole and dipole excitations. In the photonic case, we need to use core-shell systems to achieve the accidental degeneracy. Using effective medium theories, we can map these structures to a material with effective permittivity and permeability simultaneously equal to zero in the photonic system, and with effective mass density and reciprocal of bulk modulus simultaneously equal to zero in the phononic system. These systems are platforms that enable us to extend the notion of "Dirac points" from 2D to 3D. As the density of states near the 3D Dirac points are different from those near 2D Dirac points, new transport properties may emerge from those systems. These would be interesting topics for further studies. In 2D systems with Dirac cones at the zone center, only some components in the constitutive relationship are zero. In 3D, all components are zero at the Dirac point. As "zero-index" effective media, they are probably more interesting than their 2D counterparts.

## Acknowledgement

The work is supported by Hong Kong Research Grant Council GRF grant 600311, CRF grant HKUST2/CRF/11G and an internal grant SRFI11SC07.

# Appendix

In this appendix, we extend the tight-binding method introduced by Sakoda in Ref. [21] to show the existence of the 3D ADIDP in the simple cubic photonic crystal with core-shell spheres. We will use the same notations as Ref. [21]. The 3D ADIDP is a six-fold degenerate state, comprising the $T_{1u}$ and $T_{1g}$ modes. The eigen fields of the 3D ADIDP near $\Gamma$ point ($\vec{H}_k(\vec{r})$) can be described by the linear combination of the basis functions with the same symmetry as the $T_{1u}$ and $T_{1g}$ modes.

$$\vec{H}_k(\vec{r}) = \frac{1}{V} \sum_{l,m,n} e^{i\vec{k}\cdot\vec{r}_{lmn}} \sum_{i=1}^{6} A_i \vec{H}^{(i)}(\vec{r} - \vec{r}_{lmn}) \qquad (A1)$$

Here, $V$ is the volume of the unit cell, $\vec{r}_{lmn} = l\begin{pmatrix}a\\0\\0\end{pmatrix} + m\begin{pmatrix}0\\a\\0\end{pmatrix} + n\begin{pmatrix}0\\0\\a\end{pmatrix}$ is a lattice vector of the simple cubic structure, $l, m, n$ are integer. $A_i$ is the coefficient of the linear combination. $\vec{k}$ is Bloch wave vector. $\vec{H}^{(i)}(\vec{r})$ is the basis function comprising $T_{1u}$ or $T_{1g}$ modes. The indices $i=1, 2, 3$ represent the $T_{1u}$ mode, and $i=4, 5, 6$ for $T_{1g}$ mode. From the Maxwell equation, the eigenvalue equation is given by:

$$\mathcal{L}\vec{H}_k(\vec{r}) = \frac{\omega_k^2}{c^2} \vec{H}_k(\vec{r}) \qquad (A2)$$

Here, $\mathcal{L} = \nabla \times (\frac{1}{\varepsilon(\vec{r})}\nabla\times)$. $\varepsilon(\vec{r})$ is the permittivity. The eigen-frequency is given by $\omega_k$. The electromagnetic transfer integral in three dimensions can be defined by:

$$L_{lmn}^{(ij)} = \frac{1}{V} \int d\vec{r}\, \vec{H}^{(i)*}(\vec{r}) \cdot \mathcal{L}\vec{H}^{(j)}(\vec{r} - \vec{r}_{lmn}) \qquad (A3)$$

Using symmetry, we can obtain the functional form of $L_{lmn}^{(ij)}$ through Eq. (A3). In this process, we will use the symmetry operator to operate onto the eigen fields ($\vec{H}^{(i)}(\vec{r})$) of $T_{1u}$ and $T_{1g}$ modes.

The matrix representations of $T_{1u}$ mode are already shown in Ref. [21]. For ease of reference, we write down some of them again in the following.

$$\sigma_x : \begin{pmatrix} -1 & 0 & 0 \\ 0 & 1 & 0 \\ 0 & 0 & 1 \end{pmatrix}; \quad \sigma_y : \begin{pmatrix} 1 & 0 & 0 \\ 0 & -1 & 0 \\ 0 & 0 & 1 \end{pmatrix}; \quad \sigma_z : \begin{pmatrix} 1 & 0 & 0 \\ 0 & 1 & 0 \\ 0 & 0 & -1 \end{pmatrix};$$

$$C_{4x} : \begin{pmatrix} 1 & 0 & 0 \\ 0 & 0 & 1 \\ 0 & -1 & 0 \end{pmatrix}; \quad C_{4y} : \begin{pmatrix} 0 & 0 & -1 \\ 0 & 1 & 0 \\ 1 & 0 & 0 \end{pmatrix}; \quad C_{4z} : \begin{pmatrix} 0 & 1 & 0 \\ -1 & 0 & 0 \\ 0 & 0 & 1 \end{pmatrix};$$

$$C_{4x}^{-1} : \begin{pmatrix} 1 & 0 & 0 \\ 0 & 0 & -1 \\ 0 & 1 & 0 \end{pmatrix}; \quad C_{4y}^{-1} : \begin{pmatrix} 0 & 0 & 1 \\ 0 & 1 & 0 \\ -1 & 0 & 0 \end{pmatrix}; \quad C_{4z}^{-1} : \begin{pmatrix} 0 & -1 & 0 \\ 1 & 0 & 0 \\ 0 & 0 & 1 \end{pmatrix};$$

$$C_{3(111)} : \begin{pmatrix} 0 & 1 & 0 \\ 0 & 0 & 1 \\ 1 & 0 & 0 \end{pmatrix}; \quad C_{3(111)}^{-1} : \begin{pmatrix} 0 & 0 & 1 \\ 1 & 0 & 0 \\ 0 & 1 & 0 \end{pmatrix};$$

The polynomial representation of $T_{1g}$ mode is given by $\{yz(y^2-z^2), zx(z^2-x^2), xy(x^2-y^2)\}$ [38]. The matrix representations are shown below.

$$\sigma_x : \begin{pmatrix} 1 & 0 & 0 \\ 0 & -1 & 0 \\ 0 & 0 & -1 \end{pmatrix}; \quad \sigma_y : \begin{pmatrix} -1 & 0 & 0 \\ 0 & 1 & 0 \\ 0 & 0 & -1 \end{pmatrix}; \quad \sigma_z : \begin{pmatrix} -1 & 0 & 0 \\ 0 & -1 & 0 \\ 0 & 0 & 1 \end{pmatrix};$$

$$C_{4x} : \begin{pmatrix} 1 & 0 & 0 \\ 0 & 0 & 1 \\ 0 & -1 & 0 \end{pmatrix}; \quad C_{4y} : \begin{pmatrix} 0 & 0 & -1 \\ 0 & 1 & 0 \\ 1 & 0 & 0 \end{pmatrix}; \quad C_{4z} : \begin{pmatrix} 0 & 1 & 0 \\ -1 & 0 & 0 \\ 0 & 0 & 1 \end{pmatrix};$$

$$C_{4x}^{-1} : \begin{pmatrix} 1 & 0 & 0 \\ 0 & 0 & -1 \\ 0 & 1 & 0 \end{pmatrix}; \quad C_{4y}^{-1} : \begin{pmatrix} 0 & 0 & 1 \\ 0 & 1 & 0 \\ -1 & 0 & 0 \end{pmatrix}; \quad C_{4z}^{-1} : \begin{pmatrix} 0 & -1 & 0 \\ 1 & 0 & 0 \\ 0 & 0 & 1 \end{pmatrix};$$

$$C_{3(111)}: \begin{pmatrix} 0 & 1 & 0 \\ 0 & 0 & 1 \\ 1 & 0 & 0 \end{pmatrix}; \quad C_{3(111)}^{-1}: \begin{pmatrix} 0 & 0 & 1 \\ 1 & 0 & 0 \\ 0 & 1 & 0 \end{pmatrix};$$

According to Eq. (A3), we calculate $L_{lmn}^{(ij)}$ for $T_{1u}$ mode [21].

$$L_{000}^{(11)} = L_{000}^{(22)} = L_{000}^{(33)} = \frac{\omega_1^2}{c^2} + M_1$$

$$L_{\pm 1,00}^{(11)} = L_{0,\pm 1,0}^{(22)} = L_{00,\pm 1}^{(33)} = M_1'$$

$$L_{0,\pm 1,0}^{(11)} = L_{00,\pm 1}^{(11)} = L_{\pm 1,00}^{(22)} = L_{00,\pm 1}^{(22)} = L_{\pm 1,00}^{(33)} = L_{0,\pm 1,0}^{(33)} = M_1''$$

$$L_{000}^{(12)} = L_{000}^{(21)} = L_{000}^{(13)} = L_{000}^{(31)} = L_{000}^{(23)} = L_{000}^{(32)} = 0$$

$$L_{\pm 1,00}^{(12)} = L_{0,\pm 1,0}^{(12)} = L_{00,\pm 1}^{(12)} = L_{\pm 1,00}^{(21)} = L_{0,\pm 1,0}^{(21)} = L_{00,\pm 1}^{(21)} = L_{\pm 1,00}^{(13)} = L_{0,\pm 1,0}^{(13)} = L_{00,\pm 1}^{(13)} = L_{\pm 1,00}^{(31)}$$
$$= L_{0,\pm 1,0}^{(31)} = L_{00,\pm 1}^{(31)} = L_{\pm 1,00}^{(23)} = L_{0,\pm 1,0}^{(23)} = L_{00,\pm 1}^{(23)} = L_{\pm 1,00}^{(32)} = L_{0,\pm 1,0}^{(32)} = L_{00,\pm 1}^{(32)} = 0$$

Here, $\omega_1$ is the eigen frequency of $T_{1u}$ mode.

Next, we give the expression for $L_{lmn}^{(ij)}$ for $T_{1g}$ mode. The results are very similar to the $T_{1u}$ mode.

$$L_{000}^{(44)} = L_{000}^{(55)} = L_{000}^{(66)} = \frac{\omega_2^2}{c^2} + M_2$$

$$L_{\pm 1,00}^{(44)} = L_{0,\pm 1,0}^{(55)} = L_{00,\pm 1}^{(66)} = M_2'$$

$$L_{0,\pm 1,0}^{(44)} = L_{00,\pm 1}^{(44)} = L_{\pm 1,00}^{(55)} = L_{00,\pm 1}^{(55)} = L_{\pm 1,00}^{(66)} = L_{0,\pm 1,0}^{(66)} = M_2''$$

$$L_{000}^{(45)} = L_{000}^{(54)} = L_{000}^{(46)} = L_{000}^{(64)} = L_{000}^{(56)} = L_{000}^{(65)} = 0$$

$$L_{\pm 1,00}^{(45)} = L_{0,\pm 1,0}^{(45)} = L_{00,\pm 1}^{(45)} = L_{\pm 1,00}^{(54)} = L_{0,\pm 1,0}^{(54)} = L_{00,\pm 1}^{(54)} = L_{\pm 1,00}^{(46)} = L_{0,\pm 1,0}^{(46)} = L_{00,\pm 1}^{(46)} = L_{\pm 1,00}^{(64)}$$
$$= L_{0,\pm 1,0}^{(64)} = L_{00,\pm 1}^{(64)} = L_{\pm 1,00}^{(56)} = L_{0,\pm 1,0}^{(56)} = L_{00,\pm 1}^{(56)} = L_{\pm 1,00}^{(65)} = L_{0,\pm 1,0}^{(65)} = L_{00,\pm 1}^{(65)} = 0$$

Here, $\omega_2$ is the eigen frequency of $T_{1g}$ mode.

At last, we should consider the interaction terms between $T_{1u}$ and $T_{1g}$ modes. The results are shown in the following.

$$\pm L^{(15)}_{00,\pm 1} = \pm L^{(16)}_{0,\mp 1,0} = \pm L^{(24)}_{00,\mp 1} = \pm L^{(26)}_{\pm 1,00} = \pm L^{(34)}_{0,\pm 1,0} = \pm L^{(35)}_{\mp 1,00} = M_3$$

Using the relation $L^{(ij)}_{lmn} = L^{(ji)*}_{-l,-m,-n}$, we can obtain:

$$\pm L^{(51)}_{00,\mp 1} = \pm L^{(61)}_{0,\pm 1,0} = \pm L^{(42)}_{00,\pm 1} = \pm L^{(62)}_{\mp 1,00} = \pm L^{(43)}_{0,\mp 1,0} = \pm L^{(53)}_{\pm 1,00} = M_3^*$$

$M_3^*$ is the conjugate of $M_3$. The other interaction terms are equal to zero. The secular equation of Eq. (A2) is:

$$\left| \mathbf{B} - \frac{\omega_k^2}{c^2} \mathbf{I} \right| = 0. \qquad (A4)$$

Here, $\mathbf{I}$ is the unit matrix. The elements ($B_{ij}$) of $\mathbf{B}$ is defined by:

$$B_{ij} = \sum_{lmn} e^{ia(k_x l + k_y m + k_z n)} L^{(ij)}_{lmn}$$

To keep the mathematics manageable, we only consider nearest neighbor hopping. By substituting the expressions of $L^{(ij)}_{lmn}$ into the above equation, we obtain the expressions of $B_{ij}$.

$$B_{11} = \frac{\omega_1^2}{c^2} + M_1 + 2M_1' \cos(k_x a) + 2M_1'' \left[ \cos(k_y a) + \cos(k_z a) \right]$$

$$B_{15} = 2iM_3 \sin(k_z a)$$

$$B_{16} = -2iM_3 \sin(k_y a)$$

$$B_{22} = \frac{\omega_1^2}{c^2} + M_1 + 2M_1' \cos(k_y a) + 2M_1'' \left[ \cos(k_z a) + \cos(k_x a) \right]$$

$$B_{24} = -2iM_3 \sin(k_z a)$$

$$B_{26} = 2iM_3 \sin(k_x a)$$

$$B_{33} = \frac{\omega_1^2}{c^2} + M_1 + 2M_1' \cos(k_z a) + 2M_1'' \left[\cos(k_x a) + \cos(k_y a)\right]$$

$$B_{34} = 2iM_3 \sin(k_y a)$$

$$B_{35} = -2iM_3 \sin(k_x a)$$

$$B_{42} = 2iM_3^* \sin(k_z a)$$

$$B_{43} = -2iM_3^* \sin(k_y a)$$

$$B_{44} = \frac{\omega_2^2}{c^2} + M_2 + 2M_2' \cos(k_x a) + 2M_2'' \left[\cos(k_y a) + \cos(k_z a)\right]$$

$$B_{51} = -2iM_3^* \sin(k_z a)$$

$$B_{53} = 2iM_3^* \sin(k_x a)$$

$$B_{55} = \frac{\omega_2^2}{c^2} + M_2 + 2M_2' \cos(k_y a) + 2M_2'' \left[\cos(k_z a) + \cos(k_x a)\right]$$

$$B_{61} = 2iM_3^* \sin(k_y a)$$

$$B_{62} = -2iM_3^* \sin(k_x a)$$

$$B_{66} = \frac{\omega_2^2}{c^2} + M_2 + 2M_2' \cos(k_z a) + 2M_2'' \left[\cos(k_x a) + \cos(k_y a)\right]$$

$$B_{12} = B_{13} = B_{14} = B_{21} = B_{23} = B_{25} = B_{31} = B_{32} = B_{36} = B_{41}$$
$$= B_{45} = B_{46} = B_{52} = B_{54} = B_{56} = B_{63} = B_{64} = B_{65} = 0$$

Based on Eq. (A4), we use the notations

$$\xi = \frac{\omega_k^2}{c^2}, \quad \xi_1 = \frac{\omega_1^2}{c^2}, \quad \xi_2 = \frac{\omega_2^2}{c^2}$$

Eq. (A4) is reduced to:

$$\xi^6 + b_5\xi^5 + b_4\xi^4 + b_3\xi^3 + b_2\xi^2 + b_1\xi + b_0 = 0 \qquad (A5)$$

At $\Gamma$ point of the Brillouin zone, the solutions of Eq. (A5) are:

$$\xi = \begin{cases} \xi_1 + M_1 + 2M_1' + 4M_1'' = \xi_\Gamma^{(1)} \text{ (triple roots)} \\ \xi_2 + M_2 + 2M_2' + 4M_2'' = \xi_\Gamma^{(2)} \text{ (triple roots)} \end{cases}$$

If we change the form of $\xi$ to $\eta$ with the relation:

$$\eta = \xi + \frac{b_5}{6} \qquad (A6)$$

Therefore, Eq. (A5) can be changed to:

$$\eta^6 + p\eta^4 + q\eta^3 + r\eta^2 + s\eta + t = 0 \qquad (A7)$$

Here,

$$p = b_4 - \frac{5}{12}b_5^2$$

$$q = b_3 - \left(\frac{5}{54}b_5^3 + \frac{2}{3}b_5 p\right)$$

$$r = b_2 - \left(\frac{5}{432}b_5^4 + \frac{1}{6}b_5^2 p + \frac{1}{2}b_5 q\right)$$

$$s = b_1 - \left(\frac{1}{1296}b_5^5 + \frac{1}{54}b_5^3 p + \frac{1}{12}b_5^2 q + \frac{1}{3}b_5 r\right)$$

$$t = b_0 - \left(\frac{1}{46656}b_5^6 + \frac{1}{1296}b_5^4 p + \frac{1}{216}b_5^3 q + \frac{1}{36}b_5^2 r + \frac{1}{6}b_5 s\right)$$

To evaluate the $\vec{k}$ dependence of the eigen value equation around $\Gamma$ point, we expand the parameters *p, q, r, s, t* in the second, third, fourth, fifth and sixth order of $k_x, k_y, k_z$, respectively.

For the accidental degeneracy of the $T_{1u}$ and $T_{1g}$ modes at $\Gamma$ point,

$$\xi_\Gamma^{(1)} = \xi_\Gamma^{(2)} = \xi_\Gamma$$

Through very cumbersome calculations, we can obtain:

$$p = -8|M_3|^2 k^2 a^2$$

$$r = 16|M_3|^4 k^4 a^4$$

$$q = s = t = 0$$

where $k = \sqrt{k_x^2 + k_y^2 + k_z^2}$.

Since $b_5 = -6\xi_\Gamma + (M_1' + 2M_1'' + M_2' + 2M_2'')k^2 a^2$

The solutions of Eq. (A5) are:

$$\xi = \begin{cases} \xi_\Gamma + 2|M_3|ka - \dfrac{1}{6}(M_1' + 2M_1'' + M_2' + 2M_2'')k^2 a^2 & (double\ roots) \\ \xi_\Gamma - 2|M_3|ka - \dfrac{1}{6}(M_1' + 2M_1'' + M_2' + 2M_2'')k^2 a^2 & (double\ roots) \\ \xi_\Gamma - \dfrac{1}{6}(M_1' + 2M_1'' + M_2' + 2M_2'')k^2 a^2 & (double\ roots) \end{cases}$$

Since $\omega_k = c\sqrt{\xi}$, and $\omega_\Gamma = c\sqrt{\xi_\Gamma}$, we obtain the dispersion of the 3D ADIDP:

$$\omega_k = \begin{cases} \omega_\Gamma + |M_3| a c^2 k / \omega_\Gamma - \left( \frac{1}{12}\left(M_1^{'} + 2M_1^{*} + M_2^{'} + 2M_2^{*}\right)/\omega_\Gamma - \frac{1}{2\omega_\Gamma^3} c^2 |M_3|^2 \right) a^2 c^2 k^2 & (double\ roots) \\ \omega_\Gamma - |M_3| ac^2 k / \omega_\Gamma - \left( \frac{1}{12}\left(M_1^{'} + 2M_1^{"} + M_2^{'} + 2M_2^{*}\right)/\omega_\Gamma - \frac{1}{2\omega_\Gamma^3} c^2 |M_3|^2 \right) a^2 c^2 k^2 & (double\ roots) \\ \omega_\Gamma - \frac{1}{12}\left(M_1^{'} + 2M_1^{"} + M_2^{'} + 2M_2^{"}\right) a^2 c^2 k^2 / \omega_\Gamma & (double\ roots) \end{cases}$$

**Figures and Captions**

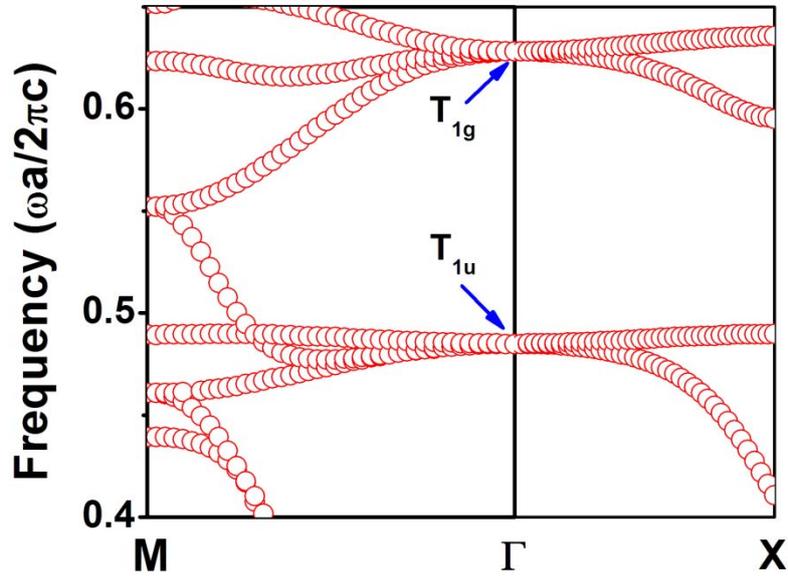

Figure 1. The photonic band structure of dielectric spheres arranged in a simple cubic lattice. The permittivity ($\varepsilon$) and permeability ($\mu$) of the sphere are $\varepsilon=12$, $\mu=1$. The radius of it is $R=0.3a$. Here, $a$ is the lattice constant. We note that the triply degenerate $T_{1g}$ and $T_{1u}$ modes have different frequencies and the dispersions are quadratic near the zone center.

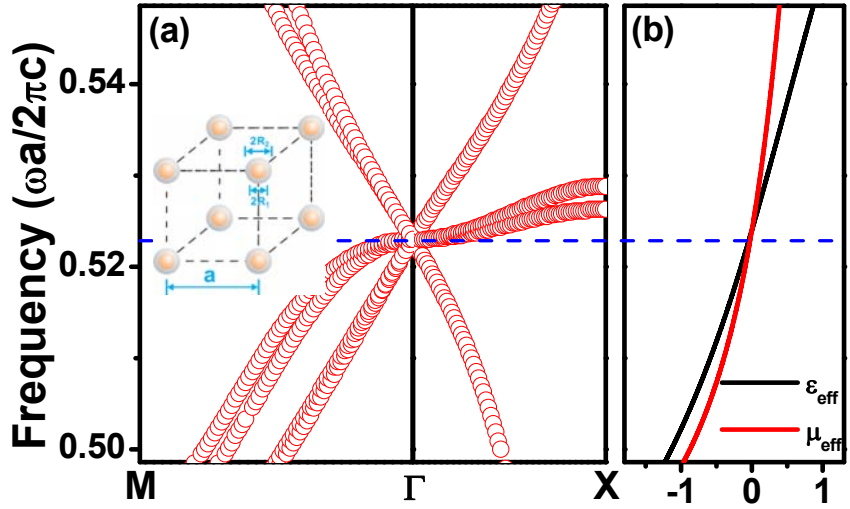

Figure 2. (a) The band structure of three dimensional core-shell photonic crystals with a simple cubic lattice. The inset is an illustration of the simple cubic unit cell. The cores (orange color) are made of perfect electric conductors, with radii of $R_1=0.102a$. The permittivity ($\varepsilon$) and permeability ($\mu$) of the dielectric shell (gray color) are $\varepsilon=12$, $\mu=1$. The radius of the outer shell is $R_2=0.3a$. Here, $a$ is the lattice constant. (b) The effective permittivity ($\varepsilon_{eff}$) and permeability ($\mu_{eff}$) as a function of frequency for this core-shell photonic crystal obtained using effective medium theory. The blue dash line marks the frequency of the Dirac point ($f_{Dirac}=0.523c/a$) in the band structure. This coincides with the frequency at which $\varepsilon_{eff}=\mu_{eff}=0$. In this case, the $T_{1g}$ and $T_{1u}$ modes are accidentally degenerate, giving rise to linear dispersions near the zone center for four of the bands.

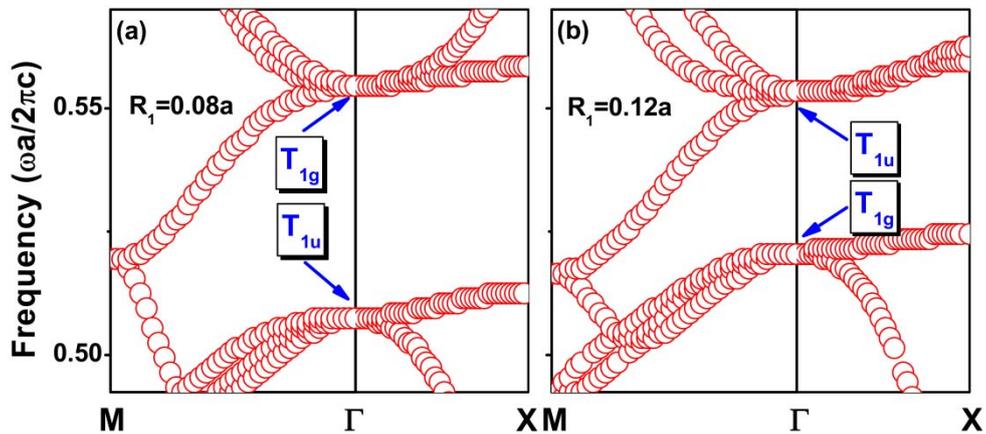

Figure 3. The band structure of three dimensional core-shell photonic crystals with simple cubic lattice for different radii of the core ($R_1$). The permittivity ($\varepsilon$) and permeability ($\mu$) of the shell are $\varepsilon=12$, $\mu=1$. The radius of it is $R_2=0.3a$. The cores are perfect electric conductors, the radii of the cores are $R_1=0.08a$ (a) and $R_1=0.12a$ (b). Here, $a$ is the lattice constant.

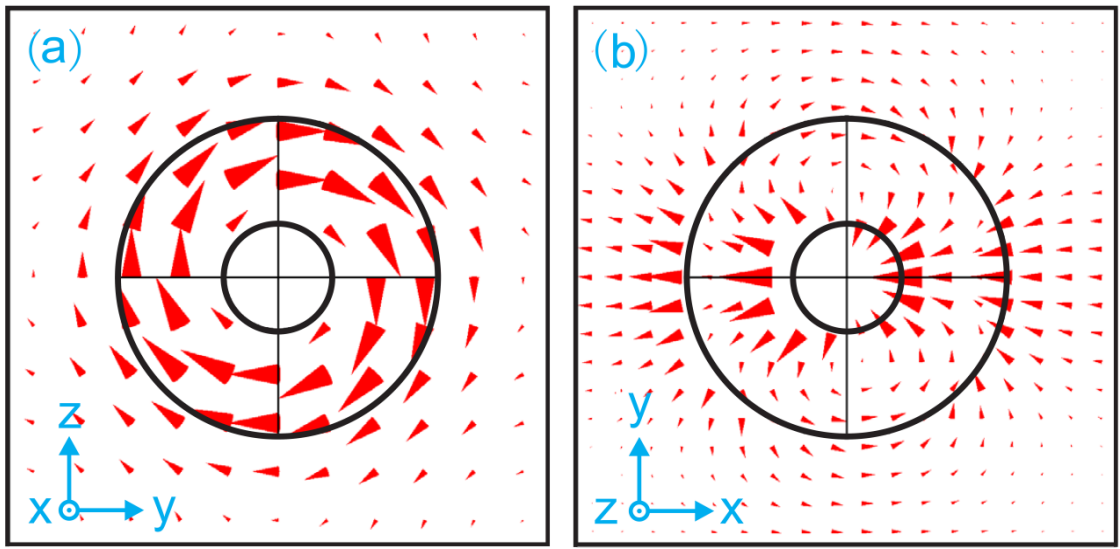

Figure 4. The electric fields of the eigenstates at the Dirac point frequency. The arrows show the directions of the electric field, the sizes of the arrows are proportional to the field amplitude. (a) The electric field in the $yz$ plane ($x=0$ plane). The field pattern shows that the eigenmode is a magnetic dipole pointing along the $x$ direction. (b) The electric field in the $xy$ plane ($z=0$ plane), implying an electric dipole excitation in $x$ direction.

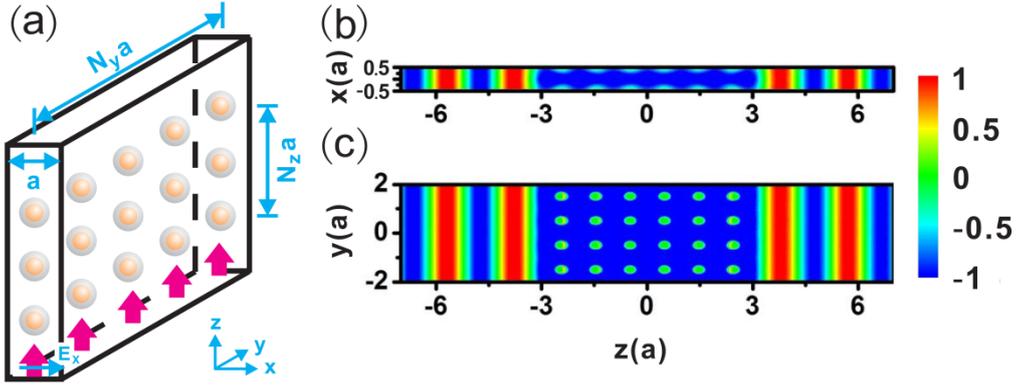

Figure 5. Finite-difference time-domain simulations showing that waves can go through the core-shell photonic crystal with very small distortions of wavefront, as expected from a $\varepsilon_{eff} = \mu_{eff} = 0$ material. (a) The super cell used for the simulations. Periodical boundary conditions are applied to the side walls (*xz*, and *yz* planes). Perfectly matched layer boundary conditions are applied at the top and bottom walls (*xy* planes). There is only one unit cell in the *x* direction and there are $N_y$ unit cells in the *y* direction. The thickness of the sample has $N_z$ unit cells. In the simulations, $N_y = 4$, $N_z = 6$. The core-shell spheres are arranged from *-1.5a* to *1.5a* and *-2.5a* to *2.5a* along *y* and *z* directions. The plane waves are incident from the negative *z* direction (as indicated by the pink arrows) with the electric field polarized along *x* direction (blue arrows). (b) the electric field pattern in the *xz* plane (*y=0* plane), and (c) the electric field pattern in the *yz* plane (*x=0* plane).

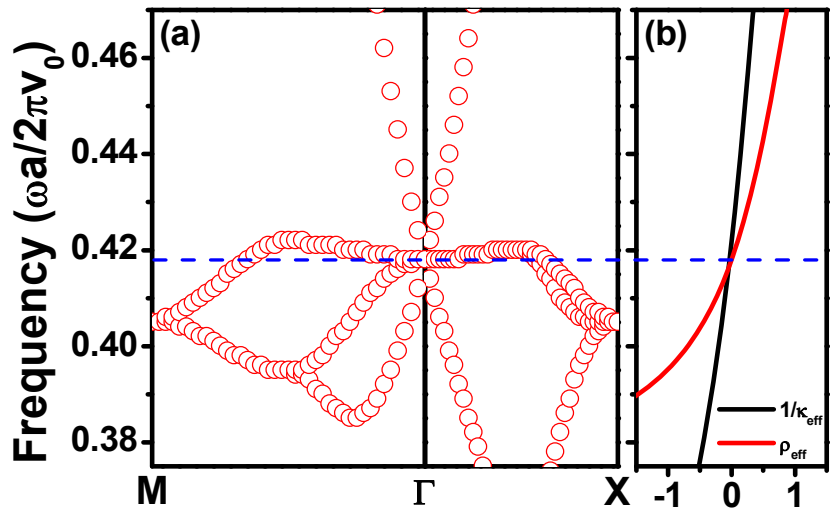

Figure 6. (a) The band structure of three dimensional simple cubic acoustic crystals consisting of rubber spheres (radius $R=0.255a$) in water. Here, $a$ is the lattice constant. The density of rubber is taken to be $\rho=1.3\times10^3 kg/m^3$, and that of water is $\rho_0=1.0\times10^3 kg/m^3$. The Lame constant in rubber is $\kappa=1.17\times10^8 N/m^2$ and for water $\kappa_0=2.22\times10^9 N/m^2$. (b) The effective mass density ($\rho_{eff}$) and reciprocal of bulk modulus ($1/\kappa_{eff}$) as a function of frequency obtained using effective medium theory for this simple cubic phononic crystal. The blue dash line marks the frequency of the Dirac point ($f_{Dirac}=0.418v_0/a$) in the band structure which coincides with the frequency at which $\rho_{eff}=1/\kappa_{eff}=0$ in the effective medium.

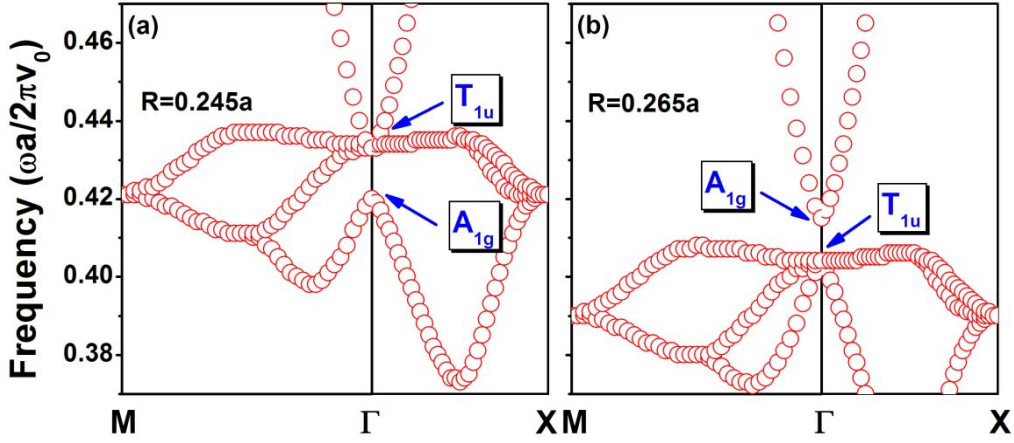

Figure 7. The band structure of three dimensional acoustic crystals arranged in a simple cubic lattice for different radii ($R$). $R=0.245a$ for Fig. 7(a), and $R=0.265a$ for Fig. 7(b). Here, $a$ is the lattice constant. The density of rubber is taken to be $\rho=1.3\times10^3 kg/m^3$, and that of water is $\rho_0=1.0\times10^3 kg/m^3$. The Lame constant in rubber is $\kappa=1.17\times10^8 N/m^2$ and for water $\kappa_0=2.22\times10^9 N/m^2$.

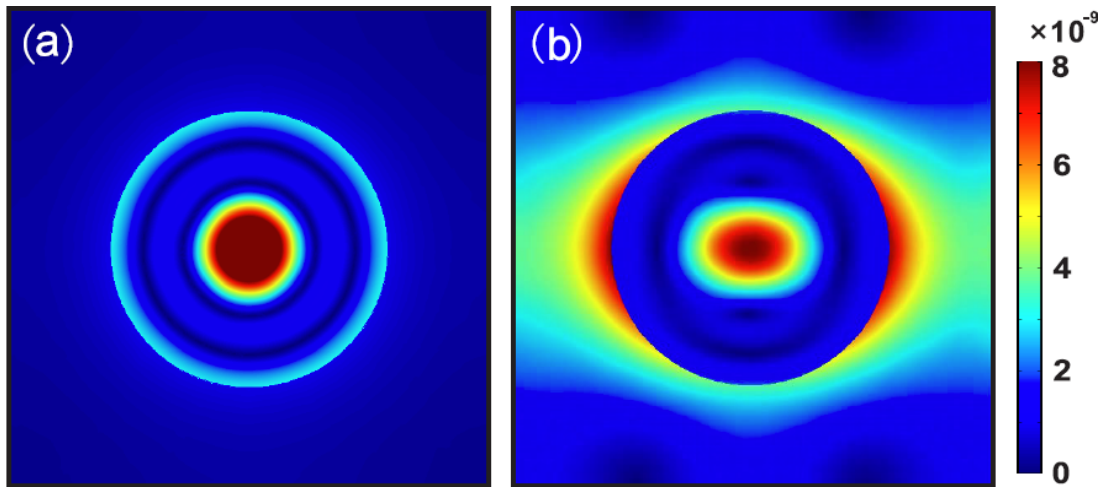

Figure 8. The displacement field patterns of the eigenstates at the Dirac point frequency ($f_{Dirac}=0.418v_0/a$). (a) implies a monopolar. (b) implies a dipolar.